\documentclass[
twocolumn,
notitlepage,
raggedbottom,
floatfix,
nofootinbib,pra]{revtex4}
\usepackage{ragged2e}
\usepackage{array}
\usepackage{comment}

\usepackage{etoolbox}
\patchcmd{\section}
  {\centering}
  {\raggedright}
  {}
  {}
\patchcmd{\subsection}
  {\centering}
  {\raggedright}
  {}
  {}
  \patchcmd{\subsubsection}
  {\centering}
  {\raggedright}
  {}
  {}

\usepackage{graphicx}
\usepackage[toc,page]{appendix}
\usepackage{bm,bbm}
\usepackage{physics}
\usepackage{amssymb}
\usepackage{amsmath}
\usepackage{dsfont}
\usepackage[usenames,dvipsnames]{xcolor}
\usepackage{soul}
\usepackage{multirow}
\usepackage{mathtools}
\usepackage{textgreek}
\usepackage{tikz}
\usetikzlibrary{cd}
\usepackage[colorlinks, linkcolor = blue, citecolor = blue, urlcolor=blue, pdfborder={0 0 0 [0 0]},bookmarks=false]{hyperref}
\usepackage[normalem]{ulem}

\usepackage{soul}

\newcommand{\R}{\mathbf{r}}
\newcommand{\G}{\mathbf{g}}
\newcommand{\Y}{\mathbf{y}}
\newcommand{\B}{\mathbf{b}}

\newcommand{\U}{\mathbf{u}}
\newcommand{\V}{\mathbf{v}}
\newcommand{\W}{\mathbf{w}}
\newcommand{\X}{\mathbf{x}}

\begin{document}

\title{Color code with a logical control-$S$ gate using transversal $T$ rotations}
\author{Benjamin J. Brown}

\affiliation{\footnotesize IBM Quantum, T. J. Watson Research Center, Yorktown Heights, New York 10598, USA}
\affiliation{\footnotesize IBM Denmark, Sundkrogsgade 11, 2100 Copenhagen, Denmark}

\begin{abstract}
The color code has been invaluable for the development of the theory of fault-tolerant logic gates using transversal rotations. Three-dimensional examples of the color code have shown us how its structure, specifically the intersection of the supports of logical operators, can give rise to non-Clifford $T$ and $CCZ$ gates. Here we present a color code with a logical control-$S$ gate that is accomplished with transversal $T$ and $T^\dagger$ rotations on its physical qubits.
\end{abstract}

\maketitle

The long-term development of fault-tolerant quantum computing is constrained by the high cost in physical resources that are needed to run large scale algorithms while the physical system also protects the encoded information from corruption using error-correction methods. To this end considerable work has been dedicated to finding ways of performing a universal set of fault-tolerant logic gates on large blocks of encoded logical qubits~\cite{KITAEV2003, Dennis02, Raussendorf_2007, Bombin_2009, Fowler12, Horsman_2012, Gottesman14, Hastings15, Brown17, Hastings18, Litinski2019game, Brown20universal, Krishna21, Cohen22,  Zhu2022, huang2022, Paletta_2024, Yamasaki:2024aa, cross2024improved, williamson2024, ide2024fault, swaroop2024universal, stein2024architecture}.  Logical gates via transversal rotations~\cite{knill1996thresholdaccuracyquantumcomputation, Bombin06, Bombin07, Bravyi12, Bombin15, Kubica15, Bombin2018transversal, Haah_2018, Vasmer19, Webster2022xpstabiliser, Quintavalle2023partitioningqubits, Webster23, Chen23, jain2024high, zhu2024, Breuckmann2024foldtransversal, wills2024constant, nguyen2024good, scruby24, lin2024transversal, golowich2024,  breuckmann2024} are desirable because they have a low time overhead and inherently minimize the propagation of errors. Discovering more general methods for performing logical operations with quantum low-density parity-check(LDPC) codes~\cite{Kovalev13, Tillich14, Hastings21, Breuckmann21balanced, Breuckmann21, Panteleev22, Leverrier2022, Bravyi:2024aa} is particularly appealing because, in addition to their the potential to encode a large number of logical qubits with a relatively few of physical qubits, we can read out their syndrome data with a low-depth quantum circuit. However, it remains difficult to find transversal operations that address specific operations on a logical code block. Pinpointing the structure and mechanisms that enable a code to perform specific logical operations will help us develop a general theory of universal fault-tolerant quantum computing.

The color code has been instrumental to the development of our theory of logic gates~\cite{Bombin06, Bombin07, Bombin07exact, Bombin15, Kubica15, bombin2018, Bombin2018transversal}. Its local construction reveals the topological structure that is required of a code to perform a logic gate at a given level of the Clifford hierarchy. Indeed, it demonstrates transversal logic gates in the highest level in the Clifford hierarchy that is achievable for a local code realized in space with fixed dimensionality~\cite{Bravyi13, Pastawski15, Jochym-OConnor18}. Furthermore, using local equivalence relations, we have used the color code to find transversal gates for the toric code~\cite{Bombin_2012, Kubica_2015, Vasmer19}, see also~\cite{Chen23, wang2023, breuckmann2024}. Logic gates in color codes on non-Euclidean lattice have also been proposed~\cite{zhu2024, scruby24}, and general algebraic constructions~\cite{Bravyi12, Haah_2018} that include the structure of the color code have also provided codes that perform non-Clifford operations on a large number of encoded qubits.

With the central role the color code has played in the discovery of transversal non-Clifford gates, it is valuable to find instances of the color code that can perform a canonical set of logical non-Clifford operations. Complementing known instances of the color code that perform $T$ and $CCZ$ gates~\cite{Bombin07, Bombin2018transversal}, in this work we demonstrate a color code that can perform a control-$S$ gate. We present the details of our results concerning the logical operators of the code at a macroscopic level. In order to do so we will state relevant rules that are derived from the microscopic details of the color code. We encourage the reader interested in these details to read Refs.~\cite{Bombin07, Bombin15, Kubica15, Brown:2016aa}. Importantly, we will frequently make use of the fact that the support of logical membranes are equivalent up to continuous deformations of their supporting manifold, i.e., up to multiplication by stabilizer operators.

The color code~\cite{Bombin07} is defined on a three-dimensional, four-valent lattice with four-colorable cells. Let us label the four colors $\R$, $\G$, $\Y$ and $\B$. We also assign colors to the faces and edges of the lattice. We say an edge that connects two cells of color $\U$ has color $\U$ and, unconventionally, we say that a face has color pair $\U \V$ if it lies at the intersection of cells of colors $\U$ and $\V$. Faces are related to edges in that the support of a face of color $\U \V$ can be composed of a disjoint set of edges of any single color $\W\not=\U, \V$. We can introduce colored boundaries to the color-code lattice. The boundary conditions determine the number of logical qubits it encodes, as well as the transversal gates that are available. A boundary of color $\U$ supports no cells of color $\U$. Later we will also introduce a Pauli-Z boundary to perform a control-$S$ gate.

\begin{figure*}
\includegraphics{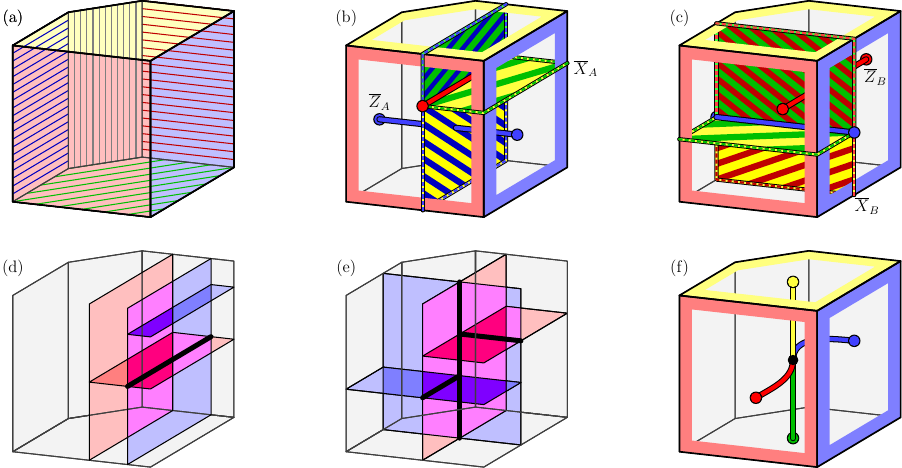}
\caption{The color code with a logical control-$S$ gate is supported on a truncated cube. (a)~The truncated cube is color coded such that exterior boundaries are shaded and interior boundaries are hatched. The cube has blue(red) boundaries on the left and right(front and back) faces. The cube also has a yellow(green) boundary on the top(bottom) of the cube. We have a Pauli-Z boundary on the truncated edge at the rear of the cube, marked with vertical grey hatching. (b)~and (c)~ show the logical operators of the color code. The membrane like supports of the logical $\overline{X}_A$ and $\overline{X}_B$ operators, and the string-like supports of the logical $\overline{Z}_A$ and $\overline{Z}_B$ operators,  are shown in~(b) and~(c), respectively. (d)~The intersection of the supports of operators $M$ and $\overline{X}_A$ is shown in black, where $M = \overline{X}_A \overline{U} \overline{X}_A \overline{U}^\dagger$  with $\overline{U}$ the transversal non-Clifford operation shown in red and $\overline{X}_A$ shown in blue. Likewise in~(e) we see the intersection of the support of $M$ and $\overline{X}_B$. (f)~The operator $\overline{X}_B M \overline{X}_B M^\dagger = i \overline{Z}_A\overline{Z}_B $.  \label{Fig:CScolorCode}}
\end{figure*}

We have two types of logical Pauli operators; string-like Pauli-Z logical operators and membrane-like Pauli-X logical operators. String operators are supported on a connected sequence of edges of the same color. The string is therefore assigned the color of the edges on which it is supported. String operators can also branch such that four strings, one of each color, meet at a point. A branching point is supported on a single additional qubit. A string of color $\U$ can terminate on a $\U$-colored boundary.  Membrane operators are composed of face operators of the same color lying on a planar  manifold. A membrane composed of faces of color $\U\V$ has color label $\U\V$. Like string operators, Membranes of appropriate colors can bifurcate along a string. Specifically, three membranes of colors $ \U\V $, $\V\W$ and $\W\U$ can bifurcate along a string of Pauli-X operators lying on edges of color $\X$ where $\U$, $\V$, $\W$ and $\X$ all take different values. A membrane operator of color $\U\V$ can terminate on a $\W$-colored boundary provided $\W \not= \U,\V$. A string-like logical operator will anti commute with a membrane-like logical operator if a string segment of color $\U$ or $\V$ pierces a membrane of color $\U\V$ an odd number of times.

Let us briefly state some important rules respected by the membrane operators. The interested reader can derive these rules from the microscopics of the color code~\cite{Bombin07, Bombin15, Kubica15}. First of all, two parallel membranes of colors $\W \U$ and $\V \W$ are equivalent to a single membrane of color $\U\V$. It is also important to look at the string-like intersections of two membrane operators that cross. The first intersection rule is that two membranes of the same color have a trivial, i.e., empty, intersection. Two membranes of colors $\U\V$ and $\U\W$ intersect along a line of color $\X$ where $\U$, $\V$, $\W$ and $\X$ are all different colors. Finally, two membranes $\U\V$ and $\W\X$ intersect along parallel strings of two different colors; either $\U $ and $\V$ or, equivalently, $\W$ and $\X$. This last rule follows from the first rule that combines two overlapping membranes together with the first intersection rule we presented.

The color code has a transversal logical operator $\overline{U} = \prod_{\textrm{even }v} T_v  \prod_{\textrm{odd }v} T_v^\dagger$, where $T_v$ is the operator $T=\textrm{diag}(1, \omega)$ acting on qubit $v$ with $\omega = e^{i\pi/4}$, and where qubits are separated into two subsets, odd and even, such that no odd qubit shares an edge with even qubit or vice versa~\cite{Bombin15, Kubica15}. The action of $\overline{U}$ depends on the boundary conditions of the color code~\cite{Bombin07, Bombin2018transversal}. We give examples of different boundary configurations in Appendix~\ref{Sec:DifferentColorCodes}.

\begin{figure*}
\includegraphics{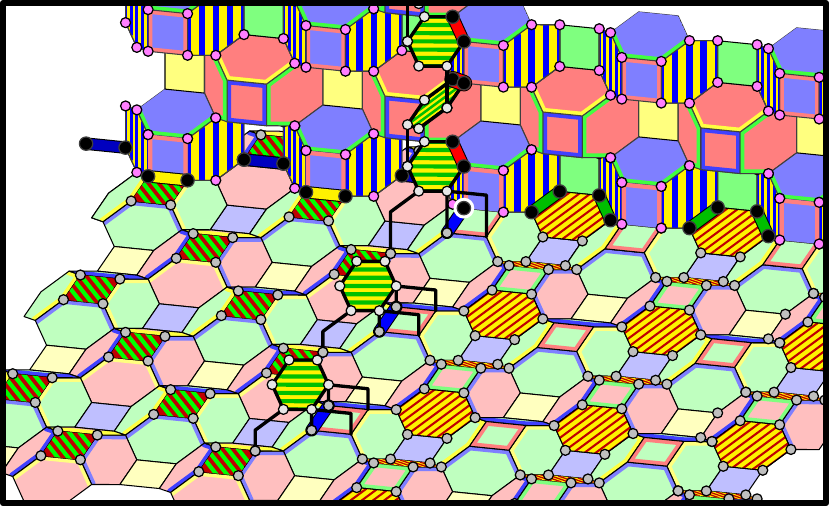}
\caption{A microscopic diagram showing the intersection between the branch of a logical Pauli-X operator $\overline{X}_B$ and a segment of the Clifford operator $M$. We show cells of a lattice geometry where we draw a $\U$-colored outline in a location where a $\U$-colored cell has been removed to reveal the operators of interest. The support of $\overline{X}_B$($M$) is shown by grey(purple) spots, and the vertices where they intersect is shown with black points. The $\overline{X}_B$ operator consists of three membranes that are made up of $\R\G$$\R\Y$, $\G\Y$ faces that meet at a blue string made up of blue edges. The $M$ operator is supported on $\Y\B$ faces on the back plane of qubits. The intersection is a branched string, where a red string composed of red edges runs up the figure,  blue and yellow strings run to the left of the figure and a green string run to the right of the figure. The branching point of the string-like intersection is marked with a white circle in the centre of the figure.
\label{Fig:Intersection}}
\end{figure*}

We will argue that the action of $\overline{U}$ on a color code on a cube with a truncated edge, as shown in Fig.~\ref{Fig:CScolorCode}(a), will give rise to a logical control-$S$ gate. See the caption of the figure for a description of the code boundaries. We derive the properties of this code in the Appendices. Precisely, we argue this code encodes two logical qubits by first considering a color code defined on a cube with only colored boundaries, and then introducing the truncated boundary by projecting the qubits in the truncated region into specific eigenstates of the Pauli-Z basis. This gives rise to a Pauli-Z boundary to the three-dimensional color code. Pauli boundaries have been considered for two-dimensional color codes in Refs.~\cite{Kesselring18, Kesselring24}, but the generalization to three dimensions is straight forward. The resulting boundary can terminate all types of string logical operators, but no membrane operators. In order to derive the properties of the color code on the truncated cube we first describe the details of a color code on a cube in Appendix~\ref{Sec:ColorCodeCube}, and then we describe the truncation operation in the unfolded picture~\cite{Bombin_2012, Kubica_2015} in Appendix~\ref{Sec:Truncation}. The logical operators of the resulting code are shown in Figs.~\ref{Fig:CScolorCode}(b) and~(c).

We can use the construction of the Pauli-Z boundary to check that $\overline{U}$ preserves the code space of the color code on the truncated cube. We first consider the color code on the cubic lattice before the truncating projection. This code has only colored boundaries. It is well understood that $\overline{U}$ preserves the code space for any such color code~\cite{Bombin07, Bombin2018transversal} due to its generalized triorthogonality conditions~\cite{Bravyi12, Haah_2018}. Now, given that the operator we use to make the projection to produce the code on the truncated cube is diagonal, it must commutes with the diagonal transversal rotation on the cubic code, we can also expect the transversal rotation $\overline{U}$ to preserve the code space of the truncated code.

Before we argue the color code gives rise to a control-$S$ gate, it may be helpful for some readers to describe the features of the color code we have described in terms of the three-dimensional toric code. The color code is locally equivalent to three copies of the toric code~\cite{Bombin_2012,Kubica_2015}. A single copy of the toric code has a single type of string logical operator and a single type of membrane logical operator, that anti-commute if the string intersects the membrane an odd number of times. Given three non-interacting copies of the toric code that we assign colors $\U = \R , \G, \B$, we have that the string(membrane)-operator of the $\U$-colored toric code is equivalent to the $\U$($\U\Y$)-colored string(membrane) of the color code. The $\Y$-colored string of the color code is equivalent to the union of all three string types of the toric code. Several works have discussed how the transversal gates of the color code are manifest in the toric code~\cite{Kubica_2015, Vasmer19, Chen23, wang2023, breuckmann2024}.

We can also define the boundaries in this unfolded picture~\cite{Kubica_2015, Kesselring18}. The toric code has a rough(smooth) boundary that can terminate its string(membrane) logical operator. A $\U = \R,\G,\B$ boundary of the color code is equivalent to a rough boundary on the $\U$-colored copy of the toric code and a smooth boundary on the other two copies. The yellow boundary of the color code is equivalent to a `fold' between the three copies of the toric code~\cite{Kubica_2015}, where the fold can terminate the union of the three string types in the toric-code picture, or a pair of membranes from two of the three copies of the toric code. The Pauli-Z type boundary is equivalent to a boundary in the toric-code picture where all three copies at the boundary are rough. We describe the unfolding picture in more detail in Appendix~\ref{SubSec:Unfolded}.

With the model and its logical operators defined, let us now argue the color code has a logical control-$S$ gate. We characterize the control-$S$ gate, $CS = \textrm{diag}(1,1,1,i)$, by its group commutation relations, $[P,Q] = PQP^{-1}Q^{-1}$, with a generating set of Pauli operators. It has a symmetrical action on two qubits, $A$ and $B$, such that 
\begin{equation}
V \equiv [X_A, CS ] = S_B CZ,
\end{equation} 
where $X_A$ is the Pauli-X operator supported on qubit $A$, $CZ = CS^2$ is the control-$Z$ gate and $S_B$ is the phase gate $S = \textrm{diag}(1,i)$ supported on qubit $B$. The diagonal $CS$ operator commutes with Pauli-Z operators, $Z_A$ and $Z_B$. To complete our argument it will be important to characterize $V$. The operator $V$ of course commutes with the diagonal string operators. We otherwise have that
\begin{equation}
[ X_A,  V ]= Z_B, \quad  [X_B,  V]=   i Z_A Z_B. \label{Eqn:V}
\end{equation}

With the control-$S$ gate characterized it remains to show that $\overline{U}$ has the appropriate action on the code space of the color code we have defined. Clearly, $\overline{U}$ commutes with the string operators as all of these operators are diagonal operators in the standard computational basis, as expected. Furthermore, we note that $\overline{X}_A$ is equivalent to $\overline{X}_B$ up to a reflection of the truncated cube and a color permutation between the red and blue boundaries. As such, we can verify the action of the transversal gate of the $\overline{X}_A$ operator without loss of generality.

We are therefore interested in the operator $  M = [\overline{U},  \overline{X}_A] $. We will show this operator has the action of $V$ on the logical space; Eqn.~(\ref{Eqn:V}). To begin, let us observe that $M$ is supported on the membrane-like support of $\overline{X}_A$. Specifically, for even(odd) qubits in the support of $\overline{X}_A$ we have $[X,T] = \sqrt{\omega} S^\dagger$($[X,T^\dagger] = \sqrt{\omega}^* S$). We therefore have
\begin{equation}
M = \theta \prod_{\substack{v\textrm{even} \\ v\in\mathcal{M}}} S^\dagger_v \prod_{\substack{v\textrm{odd} \\ v\in\mathcal{M}}} S_v,
\end{equation}
where $\mathcal{M}$ are the qubits that support $\overline{X}_A$ and we find $\theta = 1$. We obtain a trivial phase $\theta$ by looking carefully at the number of odd and even qubits in the support of $\overline{X}_A$. Indeed, the operator $\overline{X}_A$ is a set of disjoint edges, where we recall that faces can also be composed of a set of disjoint edges. As every edge includes both an even and an odd qubit, we have that every edge that makes up $\overline{X}_A$ each accumulates one $\sqrt{\omega}$ phase and one $\sqrt{\omega}^*$ phase. Overall we see that all phases cancel to give $\theta = 1$.

With the membrane-like support of $M$ established, we need only check that it has an equivalent action to $ V $ on the logical space of the code. As $M$ is diagonal in the computational basis and therefore commutes with the string logical operators, we turn to the commutation relations of $M$ with $\overline{X}_A$ and $\overline{X}_B$. We find the affirmative result by inspecting the intersection of $M$ with $\overline{X}_A $ and $\overline{X}_B$. Up to continuous deformations we show these intersections in Figs.~\ref{Fig:CScolorCode}(d) and~(e), respectively.

Let us first look at the intersection of $M$ with $\overline{X}_A$. In this case we see that a $\G\Y$ segment of $M$  intersects a $\Y\B$ segment of $\overline{X}_A$. The intersection is therefore a red string. Specifically, this red string is the support of $\overline{Z}_B$. Recalling that we have the commutation relation $[X,S] = iZ$($[X,S^\dagger ] = - i Z$) on even(odd) qubits of the intersection, we have that $[\overline{X}_A,M] = \phi \overline{Z}_B$ where $\phi $ is a phase. Once again, one can check that $\phi = 1$ because the intersection of $M$ and $\overline{X}_A$ is supported on red edges only. As such, for every $\omega$ term accumulated at an odd qubit in the intersection, we find that we have exactly one $\omega^*$ term that negates the phase at each edge. Checking this for every edge that completes the support of the intersection we find $\phi = 1$, which is consistent with Eqn.(\ref{Eqn:V}).

We can similarly check the intersection between $M$ and $\overline{X}_B$; Fig.~\ref{Fig:CScolorCode}(e). In this case we consider the intersection of several differently colored segments of the manifold. Dealing with the different segements of the intersection piecewise we arrive at the commutation relation $[\overline{X}_B, M] = \eta \overline{Z}_A\overline{Z}_B$ as shown in Fig.~\ref{Fig:CScolorCode}(f). Indeed, we can check that this operator is equivalent up to continuous deformations to $ \overline{Z}_A\overline{Z}_B$ by observing that it anticommutes with both $\overline{X}_A$ and $\overline{X}_B$. This can be seen by checking that the string operator $[\overline{X}_B, M]$ can be deformed such that its blue and red branch, respectively, intersect the blue and red component of the membrane operators $\overline{X}_A$ and $\overline{X}_B$ a single time, thereby verifying that $[\overline{X}_B, M]$ anti commutes with both $\overline{X}_A$ and $\overline{X}_B$.  A reader interested in the microscopic details of the intersection can check Fig.~\ref{Fig:Intersection} and its accompanying caption.

With regards to the phase in this commutation relation we find $\eta = \pm i$ where the sign is determined by the free choice of odd and even subsets of qubits. To see why we obtain an $i$ term, we find that the intersection of $M$ and $\overline{X}_B $ is a branching string. Each of the colored strings that emanate from the branch are composed of edges which, as we have already discussed, do not introduce a non-trivial phase to the commutator. However, the intersection of $M$ and $\overline{X}_B$ also contains a branching point supported on a single qubit. The additional branching site in the intersection of these operators therefore contributes an additional $i$ phase as expected; see Eqn.~(\ref{Eqn:V}).

To summarize, the calculations verify result, namely, that $\overline{U}$ acts like a control-$S$ gate on the color code model on the truncated cube. In the future it may be interesting to utilize our gate in fault-tolerant quantum-computing architectures based on the color code~\cite{landahl2014, Kesselring24, Thomsen2024, lee2024}. For instance, one can show that we can use a dimension jump~\cite{Raussendorf05, Bombin_2016} to project the code space of the code we have presented here onto a two-dimensional code, supported on either the yellow or green boundary. In a similar spirit the gate could be adapted for a $2+1$-dimensional implementation using just-in-time decoding~\cite{bombin2018, Brown2020, Scruby2022numerical, davydova2025universal}.

\begin{figure*}
\includegraphics{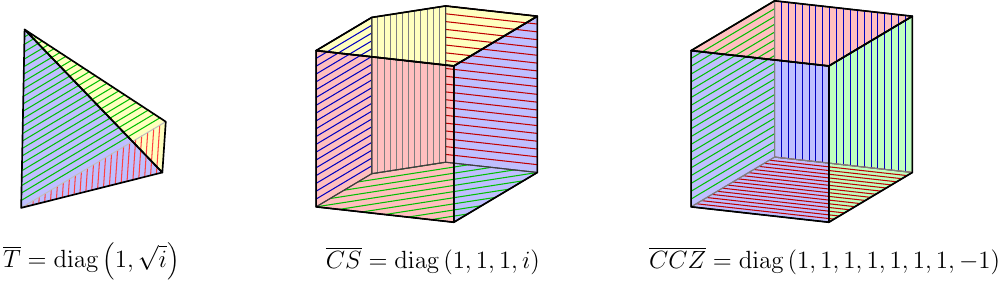}
\caption{The boundary conditions of three-dimensional color codes with different logical gates from transversal $T$ rotations. The left figure shows a color code on a tetrahedron where the transversal-$T$ rotation gives rise to a logical-$T$ gate~\cite{Bombin07}. Each face of the tetrahedron supports a distinctly colored boundary of the four possible boundary colors; red, green, yellow and blue. To the right of the figure we see a color code supported on a cube that encodes three logical qubits where a transversal-$T$ rotation gives rise to a logical controlled-controlled-$Z$ gate. The faces support boundaries of one of three colors; red, green or blue, where opposite faces of the cube support boundaries of the same color~\cite{Kubica_2015, Bombin2018transversal}. The central figure shows the color code introduced in the main text. \label{Fig:DifferentColorCodes}}
\end{figure*}

It is novel that, in contrast to other transversal gates with the color code, we have used a Pauli-Z boundary to obtain a control-$S$ gate. Looking forward, a general theory for transversal gates will be invaluable to our understanding of the fundamental limits for how quickly a fault-tolerant quantum computation can be executed. We hope that the example presented here may provide new insights towards such a theory.

\begin{acknowledgements}
I thank M. Webster for pointing out a small instance of the color code we have defined, see Fig.~6 of Ref.~\cite{Webster23}. We review this small code in the context of the exposition given in the present manuscript in Appendix~\ref{Sec:SmallCode}. I am also grateful to M. Davydova, M. Vasmer and G. Zhu for encouraging me to write up this result, and I thank the Center for Quantum Devices at the University of Copenhagen for their hospitality. Part of this research was performed while I was visiting the Institute for Mathematical and Statistical Innovation (IMSI), which is supported by the National Science Foundation (Grant No. DMS-1929348).
\end{acknowledgements}

\appendix

\section{Three-dimensional color codes}
\label{Sec:DifferentColorCodes}

The color code introduced in the main text complements known examples of color code that give rise to different diagonal logical operations upon application of a transversal-$T$ rotation on its physical qubits. The logical action is determined by the boundary conditions of the color code that undergoes the transversal rotation. In Fig.~\ref{Fig:DifferentColorCodes} we compare our example whose transversal-$T$ rotation gives rise to a logical controlled-$S$ gate to other examples of color codes.

\section{A color code defined on a cube}
\label{Sec:ColorCodeCube}

We argue that the action of $\overline{U}$ on a color code on a cube with a truncated edge, as shown in Fig.~1(a) in the main text will give rise to a logical control-$S$ gate. Let us first look at the logical degrees of freedom of this code. To begin, let us consider a cube without a truncated edge, where we have two distinct blue and red boundaries on opposite faces of the cube, together with one green and one yellow boundary lying on opposite faces, see Fig.~\ref{Fig:Untruncated}(a).

\subsection{Logical degrees of freedom}

We can characterize the logical degrees of freedom using a basis where we count the number of strings that terminate at each boundary modulo two. Let us write down an orthogonal basis of code states $ |\pm \rangle_{\mathbf{r},1}  |\pm \rangle_{\mathbf{r},2}  |\pm \rangle_{\mathbf{b},1}  |\pm \rangle_{\mathbf{b},2}  |\pm \rangle_{ \mathbf{g} } |\pm \rangle_{\mathbf{y} } $, where the sign indicates if an odd or even parity of strings terminate at the boundary that is denoted by the index of the ket vector. As there are two red and two blue boundaries, we also append numbers to their indices to distinguish them. We can measure the parity of strings terminating at a given boundary with a Pauli-X membrane operator. This operator is the product of Pauli-X terms supported on all the qubits of a given boundary. See, e.g., Refs.~\cite{Bombin07, Bombin15, Kubica15, Brown:2016aa}, for the microscopic details of these operators. For the macroscopic overview we present here, let us write down these operators as $X_B$ with indices $B = (\mathbf{r},1), (\mathbf{r},2), (\mathbf{b},1), (\mathbf{b},2), \mathbf{g}, \mathbf{y}$ labelling the different boundaries. Specifically, we have that $X_B |\pm \rangle_B = \pm |\pm \rangle_B$.

\begin{figure*}
\includegraphics{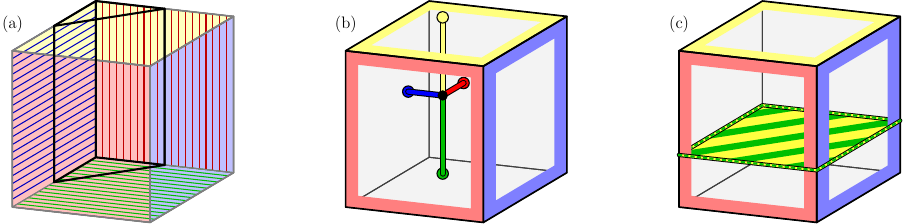}
\caption{\label{Fig:Untruncated} The color code on a cube with two red, boundaries, two blue boundaries, a green and a yellow boundary. (a)~We show the boundary configuration of the lattice on a cube. We outline the region that is truncated with dark bold lines. (b)~We show a string-like logical operator for the color code shown in~(a). The Pauli-Z string operator has a branching point where one string of each color meet. Each branch of the string operator terminates at a boundary of its respective color. This operator therefore changes the parity of strings terminating at each of these four differently colored boundaries. Importantly, this operator is completely supported in the truncated region shown in~(a). (c)~The support of a Pauli-X membrane operator that anti-commutes with the string operator shown in~(b). This membrane operator effectively counts the parity of the number of string operators that terminate at the green boundary.}
\end{figure*}

As we will see, although we can write down arbitrary basis state configurations, only certain code states are allowed. We can find the allowed states by looking at the allowed string operators in the color-code model. We will argue that the color code shown in Fig.~\ref{Fig:Untruncated}(a) has three logical qubits. To begin, let start by stating that we can readily prepare the state $|+ \rangle_{\mathbf{r},1} |+ \rangle_{\mathbf{r},2}  |+ \rangle_{\mathbf{b},1} | + \rangle_{\mathbf{b},2}  | + \rangle_{\mathbf{g}}  | + \rangle_{\mathbf{y}} $. At the microscopic level this is achieved by preparing all the physical qubits of the code in the $|+\rangle_q$ state, and then projecting onto the $+1$ eigenvalue eigenstate of the Pauli-Z face stabilizers. This inherently initializes all Pauli-X membrane operators in their $+1$ eigenvalue eigenstate, i.e. $X_B = +1$ for all values of $B$.

We can change the values of a boundary by applying string operators that respect the branching rules defined in the main text. Let us briefly elaborate on these constraints. To recall, the color code is such that a string operator has a single color, $\mathbf{u} = \mathbf{r}  ,\mathbf{g} ,\mathbf{y}  ,\mathbf{b} $, that cannot be changed. A string of color $\mathbf{u}$ can terminate at a boundary of color $\mathbf{u}$. A string can also branch, such that a string of each color, $ \mathbf{r} $, $\mathbf{g}$, $\mathbf{y} $, and $\mathbf{b}$ extend from the branching point. These string operators can then go on to terminate at a boundary of their respective color, or meet another branching point. As a consequence of these branching rules, the total parity of the number of strings of color $\mathbf{u}$ collectively terminating at all boundaries of their respective color must be equal for all $\mathbf{u}$. In the remainder of this section we will verify that we can encode three logical qubits in the code space of the color code we have defined on a cube. Afterwards, as an aside, we will verify these rules at the level of the stabilizer group of the color code with colored boundaries.

Having recalled the allowed string logical operators let us now look at the code states that can be obtained. We first consider a string operator with a branch, see Fig.~\ref{Fig:Untruncated}(b). This operator will change the parity of four differently colored boundaries in unison. As there is only one green boundary and one yellow boundary, their value is always correlated. On the other hand we have a choice between two red boundaries and a choice between two blue boundaries. Without loss of generality let us choose this branching string such that it terminates at boundaries $(\mathbf{r},1)$ and $(\mathbf{b},1) $ together with the $\mathbf{g}$ and $\mathbf{y}$ boundaries. We show this operator in Fig.~\ref{Fig:Untruncated}(b) that has the action $Z_{\mathbf{r}, 1} Z_{\mathbf{b}, 1} Z_{\mathbf{g}} Z_{\mathbf{y}}$ on code states written in the basis defined above, whereby $Z_B |\pm\rangle_B = |\mp\rangle_B$. This specifies one degree of freedom, where this string operator anti commutes with the logical operator $X_\mathbf{g}$, shown in Fig.~\ref{Fig:Untruncated}(c), where the membrane has been continuously manipulated to leave a small amount of space between the green boundary and the membrane operator.

Our freedom to choose to terminate a string at alternative red and blue boundaries gives us two additional degrees of freedom. Indeed, we can change the parity of strings terminating at the distinct red and blue boundaries in unison. This is achieved with continuous red(blue) strings that extend between the red(blue) boundaries. Such operators are shown in the main text in Figs.~1(b) and~(c), on the truncated code. However, as these operators have no common support in the truncated region, we can also consider the action of these operators on the cubic color code before the truncation is made. These operators have the effective action $Z_{\mathbf{b},1} Z_{\mathbf{b},2}$ and $Z_{\mathbf{r},1} Z_{\mathbf{r},2}$, respectively, in the basis defined above. These operators anti-commute with membrane operators $X_{\mathbf{b},2}$, and $X_{\mathbf{r},2}$, respectively. These membrane operators are also shown in Figs.~1(b) and~(c), respectively, where these membrane operators are manipulated such that they are supported a small distance away from their respective boundaries. The fact that these membranes have a branch such that remains supported at the respective boundaries $(\mathbf{r},2)$ and $(\mathbf{b},2)$ is a feature of the boundary conditions chosen for this code. We will elucidate this feature in the unfolded picture later in Appendix~\ref{SubSec:Unfolded}.

We can therefore write down a canonical basis of logical operators for the color code shown in Fig.~\ref{Fig:Untruncated}(a) as follows:
$$
\overline{X}_A = X_{\B,2}, \quad \overline{Z}_A = Z_{\B,1} Z_{\B,2}, 
$$
\begin{equation}
\overline{X}_B = X_{\R,2}, \quad \overline{Z}_B = Z_{\R,1}Z_{\R,2}, 
\end{equation}
$$
\overline{X}_C = X_{\mathbf{g}}, \quad \overline{Z}_C = Z_{\mathbf{r},1}Z_{\mathbf{b},1} Z_{\mathbf{g}} Z_{\mathbf{y}}. 
$$
Given that, by observation, we cannot make any other string operators that respect the branching rules of the color code that change the boundary configurations in a way that cannot be generated by the application of operators $\overline{Z}_1$, $\overline{Z}_2$ or $\overline{Z}_3$. We therefore argue we have fully characterized the code space of the color code on a cube. 

This code is tri-orthogonal~\cite{Bravyi12, Bombin15} and therefore supports a non-Clifford logical operator upon the application of a transversal-$T$ rotation on its physical qubits. We leave it as an exercise to the reader to check that this action is $\overline{U} = \overline{CCZ}_{ABC} \overline{CS}_{AB}$. One can check this by following the analysis method used in the main text.

\begin{figure*}
\includegraphics{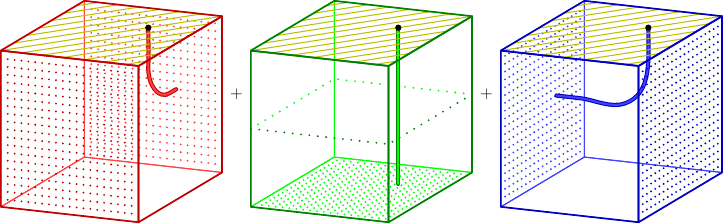}
\caption{ \label{Fig:UnfoldedLogicalZ} A string logical operator for the color code on a cubic manifold in the unfolded picture, equivalent to that shown in Fig.~\ref{Fig:Untruncated}(b). The figure shows three copies of the three-dimensional toric code; a red, green and blue copy. 
Rough boundaries are marked with small dots otherwise, with the exception of the hatched boundary at the top of each copy, all other boundaries are smooth.
The string-like logical operator terminates at rough boundaries on each of the three distinctly colored copies. The three copies are joined only at the yellow boundary. Here, string configurations that are allowed are such that a string from each of the three copies must meet at a common point. The union of three strings at this boundary is equivalent to a yellow string in the color code. As such, in the color-code picture, we have that this string-like logical operator is equivalent to that shown in Fig.~\ref{Fig:Untruncated}(b). The string logical operator anti commutes with a membrane logical operator supported on the green copy of the toric code only. A dotted line shows where the membrane-like logical operator terminates on the smooth boundaries. This membrane is equivalent to that shown in Fig.~\ref{Fig:Untruncated}(c).}
\end{figure*}

\subsection{Verifying the branching rules of string operators using the stabilizer group of the color code}

We identify the constraints the string operators must respect at the level of the stabilizer group of the code. Stabilizer operators $S$ give an Abelian subgroup of Pauli operators that specify code states of a code whereby all code states are in the $+1$ eigenvalue eigenspace of stabilizers, i.e., $S = +1$. In general, a color code with colored boundaries has stabilizer operators
\begin{equation}
S_{\mathbf{u}, \mathbf{v} } \doteq  \prod_{j} X_{(\mathbf{u}, j)} \prod_{j} X_{(\mathbf{v}, j)} , \label{Eqn:Constraint}
\end{equation}
for two distinct colors $\mathbf{u}$ and $\mathbf{v}$ where we take the product over logical boundary operators $X_B$, with indices $j$ distinguishing distinct boundaries of the same color. We explain how to obtain this result shortly. Given that $S_{\mathbf{u}, \mathbf{v}} = +1 $, we have that $\prod_{j} X_{(\mathbf{u}, j)} \prod_{j} X_{(\mathbf{v}, j)} = +1$ must have even parity or, equivalently, we may say that $\prod_{j} X_{(\mathbf{u}, j)} =  \prod_{j} X_{(\mathbf{v}, j)} $ for any color pair. This constrains the number of string operators that can terminate at the collective set of boundaries of the same color to be equal for any pair of colors. This constraint is respected by the allowed string operators defined in the main text.

We check that operators $S_{\mathbf{u}, \mathbf{v} } $ are elements of the stabilizer group as follows, thereby verifying Eqn.~(\ref{Eqn:Constraint}). With the common definition of the color code and its colored boundaries~\cite{Bombin07, Brown:2016aa}, its stabilizer group has four types of Pauli-X cell operators $S_c = +1$ where each cell has one of four colors, $\mathbf{col}(c) = \mathbf{r}, \mathbf{g}, \mathbf{y},\mathbf{b}$. One can check that the product of all cells of any two colors is such that:
\begin{equation}
 \prod_{c: \mathbf{col}(c) = \mathbf{u}, \mathbf{v}} S_c = \prod_{j} X_{(\mathbf{u}, j)} \prod_{j} X_{(\mathbf{v}, j)} = S_{\mathbf{u}, \mathbf{v}},
\end{equation}
thereby verifying the allowed string rules for the color code.

\section{Truncating the color code on a cube}

\label{Sec:Truncation}

With the color code defined on a cube with colored boundaries, we can now describe the properties of the truncated color code discussed in the main text by truncating an edge of the color code on a cube via projection. To understand the mechanics of the truncation operation, it is first helpful to describe the color code in the unfolded picture.

\subsection{Unfolding the color code}

\label{SubSec:Unfolded}

Let us briefly describe the model defined in the previous Section in the unfolded picture. The three-dimensional color code is equivalent to three copies of the three-dimensional toric code up to a local constant depth unitary circuit~\cite{Kubica_2015}. As such, in the toric-code picture, we can find a set of logical operators with equivalent action and similar support as in the color code picture, up to the small light cone of the constant depth circuit.

\begin{figure*}
\includegraphics{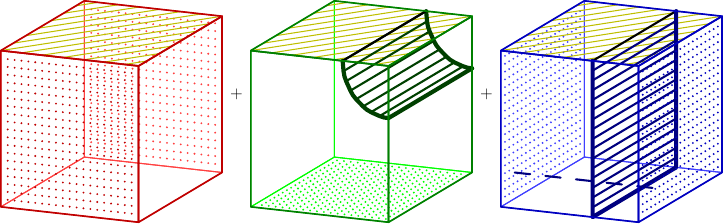}
\caption{ \label{Fig:UnfoldedLogicalX} A membrane logical operator shown in the unfolded picture equivalent to that shown in Fig.~1(b) in the main text. The membrane is supported on the blue and green copies of the toric code. At the hatched boundary at the top of the manifold where the three copies meet, membrane logical operators from two of the three copies must share a common line. The meeting point is shown by a thick black line. This membrane operator anti commutes with a string-like logical operator supported on the blue copy of the toric code. This string operator that intersects the membrane is marked by a dashed line.}
\end{figure*}

Let us briefly state the key results of the unfolding work at a macroscopic level, concentrating on the boundaries and logical operators in the unfolded picture. We leave the reader to see Ref.~\cite{Kubica_2015} to learn the derivation of these results. We show the the color code described above, as in Fig.~\ref{Fig:Untruncated}(a), in its unfolded picture in Figs.~\ref{Fig:UnfoldedLogicalZ} and~\ref{Fig:UnfoldedLogicalX}. While the unfolding circuit leaves the three copies of the toric code overlapping, in the figures we show the three copies spaced out side by side for clarity.

A single copy of the three-dimensional toric code has two boundary types; a rough boundary and a smooth boundary. The Rough boundary terminates string-like Pauli-Z logical operators and a smooth boundary terminates a Pauli-X membrane operator. A Pauli-X membrane operator anti-commutes with a Pauli-Z string operator if the string intersects the membrane an odd number of times. Under the unfolding, we distinguish the three copies of the toric code by color; $\mathbf{r}$, $\mathbf{g}$ and $\mathbf{b}$. String and membrane logical operators in the different copies of the toric code correspond to differently colored logical operators in the color code.

Let us summarize the equivalent logical operators. A string operator appearing in the $\mathbf{u} = \mathbf{r}, \mathbf{g}, \mathbf{b}$ colored toric code corresponds to a $\mathbf{u}$ colored string in the color-code model. The yellow-string in the color-code model corresponds to the union of strings in all three copies of the toric code running together along the same path.

A membrane operator in the $\mathbf{u} = \mathbf{r}, \mathbf{g}, \mathbf{b}$ colored toric code is equivalent to the $\mathbf{uy}$-colored membrane in the color code model. In the color-code picture we have that the combination of two membranes of color $\mathbf{uy}$ and $\mathbf{vy}$ is equivalent to a single membrane of color $\mathbf{uv}$. Equivalently then, the combination of two membranes in the toric code picture of colors $\mathbf{u}$ and $\mathbf{v}$ are equivalent to a color code membrane of color $\mathbf{uv}$ for $\mathbf{u}\not= \mathbf{v}$ and $\mathbf{u},\mathbf{v}\not= \mathbf{y}$.

To completely characterize the unfolded picture, we need to discuss how the logical operators interact with the boundaries. In Figs.~\ref{Fig:UnfoldedLogicalZ} and~\ref{Fig:UnfoldedLogicalX} we show logical operators in the toric-code picture that are equivalent to those shown in the color-code picture in Fig.~1(b), and Figs.~\ref{Fig:Untruncated}(b) and~(c), respectively. A color-code boundary of color $\mathbf{u} = \mathbf{r}, \mathbf{g}, \mathbf{b}$ is equivalent to a boundary in the toric code where the $\mathbf{u}$-colored copy of the toric code is rough and the boundaries of the other two copies are smooth. This produces a boundary that can terminate a $\mathbf{u}$-colored string or a membrane of any other color. This is equivalent to a color code boundary that can terminate a $\mathbf{u} = \mathbf{r}, \mathbf{g}, \mathbf{b}$ colored string and a $\mathbf{vy}$-colored membrane for $\mathbf{v} \not= \mathbf{u}, \mathbf{y}$. 

The yellow boundary of the color code maps non-trivially in the toric code picture, where the three copies are `folded' along this plane. At the yellow boundary in the toric-code picture we observe a branching point, where the union of strings of all three colors can terminate. We show a branching point in Fig.~\ref{Fig:UnfoldedLogicalZ} with a black point on the yellow hatched boundary of the three copies. This is equivalent to a yellow string in the color-code picture, as the yellow string is equivalent to the union of red, green and blue string types. No other strings can terminate here.

Lastly, in the color-code picture, the yellow boundary can terminate a membrane of color $\mathbf{uv}$ with $\mathbf{u},\mathbf{v}\not= \mathbf{y}$. In the unfolded picture this is equivalent to the union of membranes from two of the three copies of the toric code. As such, the join at the hatched boundary is such that a pair of membrane operators can terminate along a common line. We show one such termination in Fig.~\ref{Fig:UnfoldedLogicalX} where the union of the green and blue membrane meet at the yellow hatched boundary along a common black line.

We can therefore describe the logical operators of the color code defined on the cube in the unfolded picture. Fig.~\ref{Fig:UnfoldedLogicalZ} shows a string-like operator that terminates at one boundary of each color. In the toric-code picture the string terminates at one rough boundary of each of the three toric code copies, and it has a single branch terminating at the yellow boundary. This is equivalent to the logical operator shown in Fig.~\ref{Fig:Untruncated}(b). It anti-commutes with a membrane operator that terminates on the smooth boundaries of the green copy of the toric code. We show the terminal line of this membrane by a dotted line on the smooth boundaries of the green copy of the code. This is equivalent to the logical operator shown in Fig.~\ref{Fig:Untruncated}(c).

Likewise, we show the unfolded logical membrane operator equivalent to that in the main text, Fig.~1(b) in Fig.~\ref{Fig:UnfoldedLogicalX}. Here, two membranes, one on the green copy of the toric code and the other at the blue copy of the toric code, meet along a black line on the hatched boundary, thereby respecting the rules of the fold. This membrane, the union of a green and blue membrane in the toric-code picture, is equivalent to a $\mathbf{gb}$ membrane operator in the color-code picture. This boundary is indeed allowed to terminate at a yellow boundary. This logical operator anti-commutes with a string operator supported on the blue copy of the toric code. Its support is shown in the figure by a dashed line. The logical string operator supported by this dashed line is equivalent to the blue string logical operator shown in Fig.~1(b).

\subsection{Truncating an edge of the color code on a cube}

\begin{figure*}
\includegraphics{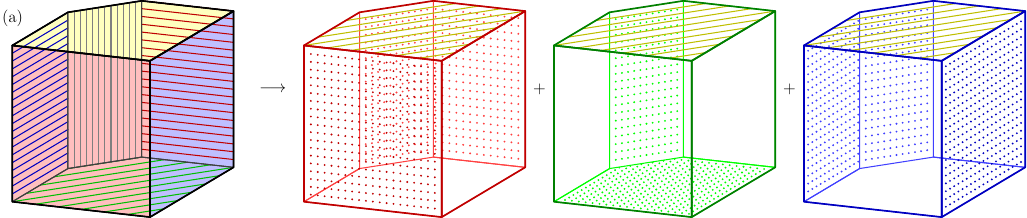}
\caption{\label{Fig:UnfoldedTruncated} The truncated color code shown in the unfolded picture. On the truncated boundary, each copy of the surface code has a rough boundary. This enables the boundary to terminate a string operator of any color or, indeed, any combination of colored strings. Furthermore, one can see that producing the boundary on the truncated face simply extends the rough boundaries of the red, green and blue copies of the toric code over the truncated edge. As such, we do not create any new modes where a string operator can terminate.}
\end{figure*}

The inclusion of a Pauli-Z boundary on a truncated edge of the cubic color code we have just described, shown in Fig.~\ref{Fig:Untruncated}(a), eliminates one logical degree of freedom. A Pauli-Z boundary can terminate a logical string of any color. However, it cannot terminate membranes of any color. Pauli boundaries are described for two-dimensional color codes in Refs.~\cite{Kesselring18, Kesselring24}. We can produce the Pauli-Z boundary by projecting all of the qubits of a region of a code onto the $+1$ eigenvalue eigenstate of the Pauli-Z operator, i.e., the zero state. The border of the projected region gives rise to a Pauli-Z boundary.

The projection on the truncated region modifies the stabilizer group of the original code defined on a cube. The operation drives all qubits $q$ on the truncated region into the $Z_q = +1$ eigenvalue eigenstate. Stabilizers or the original code that commute with this projection remain in the stabilizer group of the truncated code, whereas stabilizers that anti-commute with the projected region are removed from the stabilizer group in the new code. For the color code, Pauli-Z face terms all commute with the projection. These operators therefore remain in the stabilizer group of the truncated code. On the other hand, Pauli-X cell stabilizers with support on the truncated region will be removed from the stabilizer group. The removal of these stabilizers changes the properties of the code. Indeed, in the truncated code, string operators that terminate in the truncated region, that may have previously anti-commuted with these cell operators, now commute with the stabilizer group. Many of these operators will be Pauli-Z stabilizers of the new code, but one should be concerned that we may have introduced new logical degrees of freedom to the truncated code. In the remainder of this section let us discuss the properties of this new code. First of all, we show that one logical degree of freedom is removed from the original code defined on the cube. We will also appeal to the unfolded picture to argue that no new degrees of freedom are introduced to the code.

This projection operation removes one logical degree of freedom. One string-like logical operator of the color code on the cube can be supported entirely on the truncated region. We show the Pauli-Z logical operator next to the truncated region in Figs.~\ref{Fig:Untruncated}(a) and (b). The projection drives all of the qubits $q$ in the truncated region into the state with $Z_q = +1$. Given that the value logical string operator can be inferred from the state of the qubits that have been projected into the $Z_q = +1$ state, we also project the logical qubit $C$ into a logical eigenstate in the Pauli-Z basis $\overline{Z}_C = +1$. As such we have eliminated one logical degree of freedom.

The unfolded picture helps to reveal the properties of the code after the projection has been made. Like in the color code picture, in the unfolded picture a boundary equivalent to the Pauli-Z boundary is produced by projecting all of the qubits of a region into the $Z_q = +1$ state. It is well known in the toric-code picture that projecting a region of its qubits into an eigenstate of their Pauli-Z basis creates a rough boundary on at the border of the region where the projection is made~\cite{Raussendorf_2007, Brown17, Bombin_2009, Kesselring18, Kesselring24}. We can therefore show the Pauli-Z boundary in the unfolded picture; see Fig.~\ref{Fig:UnfoldedTruncated}. The figure shows that on each of the three colored copies of the toric code we have a rough boundary on the truncated edge of the cube. This is able to terminate Pauli-Z string-like logical operators, but no membrane operators can terminate at this boundary.

\begin{figure*}
\includegraphics{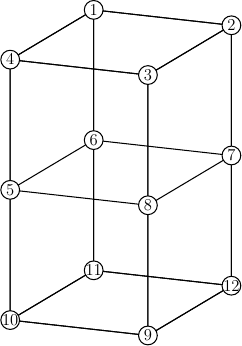}
\caption{Small instance of the color code with a logical control-$S$ gate found in Ref.~\cite{Webster23}; see Fig.~6. The code is defined on a cuboid that is divided into two cubic cells. It has Pauli-X stabilizers supported on cells; $S^X_\G = X_1 X_2 X_3X_4X_5X_6X_7X_8$ and  $S^X_\Y = X_5X_6X_7X_8 X_9 X_10 X_11 X_12$. If we interpret these cubic cells as a yellow and green colored cells, respectively, we can interpret this code as a small color code with boundaries consistent to the macroscopic codes we have presented in the main text, whereby a boundary of a given color supports no cells of its own color. As such we can regard the top(bottom) boundary as a yellow(green) boundary, and the sides of the cuboid as red and blue boundaries, where the color of these side boundaries alternate between red and blue. Like the color code, this code also has Pauli-Z type boundaries on the faces of these cubes. We can generate the face operators with Pauli-Z stabilizers $S^Z_1 = Z_1Z_2Z_3Z_4$, $S^Z_2 = Z_1Z_2Z_6Z_7$, $S^Z_3 = Z_2Z_3Z_7Z_8$,   $S^Z_4 = Z_3Z_4Z_5Z_8$, $S^Z_5 = Z_5Z_6Z_{10}Z_{11}$, $S^Z_6 = Z_6Z_7Z_{11}Z_{12}$ and $S^Z_7 = Z_7Z_8Z_9Z_{12}$. Lastly the code has a stabilizer $S^Z_0 = Z_1 Z_6 Z_{11}$. This is a special logical operator in the form of a string that extends between the top and bottom boundaries of the cuboid. We discuss this operator in the main text as, for this small case, we can view this stabilizer as the result of the truncation step we have described in Appendix~\ref{Sec:Truncation}. We can write logical operators for this code as follows: $\overline{X}_A =  X_2 X_3 X_7 X_8X_8X_12$, $\overline{Z}_A = Z_6Z_7$,  $\overline{X}_B =  X_3 X_4 X_5 X_8X_9X_10$ and $\overline{Z}_B = Z_7Z_8$. These logical operators correspond to the equivalently labeled logical operators in the main text shown in Fig.~1(b) and~(c). We leave it as an exercise to the reader to check that the operator $\overline{T} = \prod_{\textrm{odd }j} T_j \prod_{\textrm{even }j} T^\dagger_j$ acts like a logical control-$S$ gate between the two logical qubits. \label{Fig:SmallCode}}
\end{figure*}

As a final remark, as mentioned, one might be concerned that introducing a new boundary to the color code manifold might introduce new degrees of freedom. We have argued that distinct boundaries give rise to new modes where string-like logical operators can terminate. In the unfolded picture, these distinct boundaries are of rough type. The unfolded picture reveals that no new modes have been created. For all three copies of the toric code, the rough boundary on the truncated edge can be viewed as an extension of another nearby rough boundary. As such, a string terminating at a boundary adjacent to the Pauli-Z boundary can be continuously deformed such that it terminates on the Pauli-Z boundary itself. We therefore argue that no new logical qubits have been produced.

\section{A small instance of thecolor code on a truncated cube}

\label{Sec:SmallCode}
In Fig.~\ref{Fig:SmallCode} we show a small instance of the color code we propose in the main text. This small code was presented in Ref.~\cite{Webster23}. The stabilizers of the code, and its logical operators, are written down in the figure caption. Let us briefly elaborate on the truncation step for this code. In the main text we consider truncating an edge of the cubic lattice. As we have explained in Appendix~\ref{Sec:Truncation}, this effectively projects a third logical operator of the color code defined on a cube, see Appendix.~\ref{Sec:ColorCodeCube}, into an eigenstate in the computational basis, i.e., $\overline{Z}_C = +1$. We could recover a small instance of this cubic color code from the small code shown in Fig.~\ref{Fig:SmallCode} by regarding the stabilizer $S^Z_0$ as a logical operator $\overline{Z}_C = S^Z_0$. This logical operator anti-commutes with logical operator $\overline{X}_C = X_1X_2X_3X_4$. These operators correspond to those shown in Figs.~\ref{Fig:Untruncated}(b) and~(c), respectively. Including this logical operator as a stabilizer, $S^Z_0 = \overline{Z}_C = +1$, as described in the caption of Fig.~\ref{Fig:SmallCode}, effectively projects this logical operator into a computational basis state for this small instance of the code. This inclusion is analogous to the truncation step proposed for larger codes in Appendix~\ref{Sec:Truncation}.

\end{document}